\documentclass{pic2012}
\def\runauthor{E.W. Varnes}
\def\shorttitle{Tevatron Higgs Searches}
\begin{document}

\title{HIGGS BOSON SEARCHES AT THE TEVATRON
}

\author{E.W. Varnes}

\address{University of Arizona\\
1118 E. 4th St, Tucson AZ 85721, USA\\
E-mail: varnes@physics.arizona.edu }

\maketitle

\abstracts{The search for the Higgs boson, both in the context of the standard model and extensions to it, has been a key focus during Run II of the Tevatron.  I report on the status of these searches, which are highlighted by evidence at the 3 standard deviation level for the SM Higgs in its $b\bar{b}$ decay mode, the strongest direct evidence to date for fermionic couplings of the Higgs boson.
}

\section{Introduction} 

The Higgs mechanism~\cite{higgs_theory} is one of the cornerstones of the standard model (SM)~\cite{SM}. While the SM has passed many experimental tests since its formulation, direct detection of the physical Higgs boson predicted by the Higgs mechanism proved elusive.  In the period between the shutdown of LEP2 and the advent of the LHC, the Tevatron $p\bar{p}$ collider was the only facility potentially capable of producing Higgs bosons.  This report summarizes the results of the Higgs boson search at the Tevatron, using data collected by both the CDF and D0 experiments.  Some of these results are now superseded with the discovery of a Higgs-like boson at the LHC~\cite{Higgs_discovery}, but some (in particular the evidence for the $b\bar{b}$ decay mode of this new particle) remain highly relevant.

\section{General features of Tevatron searches}
Statistics are at a premium when searching for the Higgs boson at the Tevatron, so every effort is made to extract maximal information from the data.  This generally means that multivariate discriminants are used to distinguish signal from background, with boosted decision trees and neural networks being the most commonly-used.   When setting limits on the Higgs production cross section, SM branching ratios are assumed and the modified frequentist method is used.

\section{Searches for the SM Higgs boson}

Since the mass of the Higgs boson ($m_H$) is not predicted, the search program at the Tevatron is structured for sensitivity from the LEP2 limit of 114 GeV up to the end of the Tevatron's kinematic reach, $\approx 200$ GeV, using different production and decay modes in the various mass ranges.  

\subsection{Searches optimized for high masses}

\begin{figure}[!thb]
%\vspace*{4.0cm}
\begin{center}
\centerline{\epsfxsize=2.5in\epsfbox{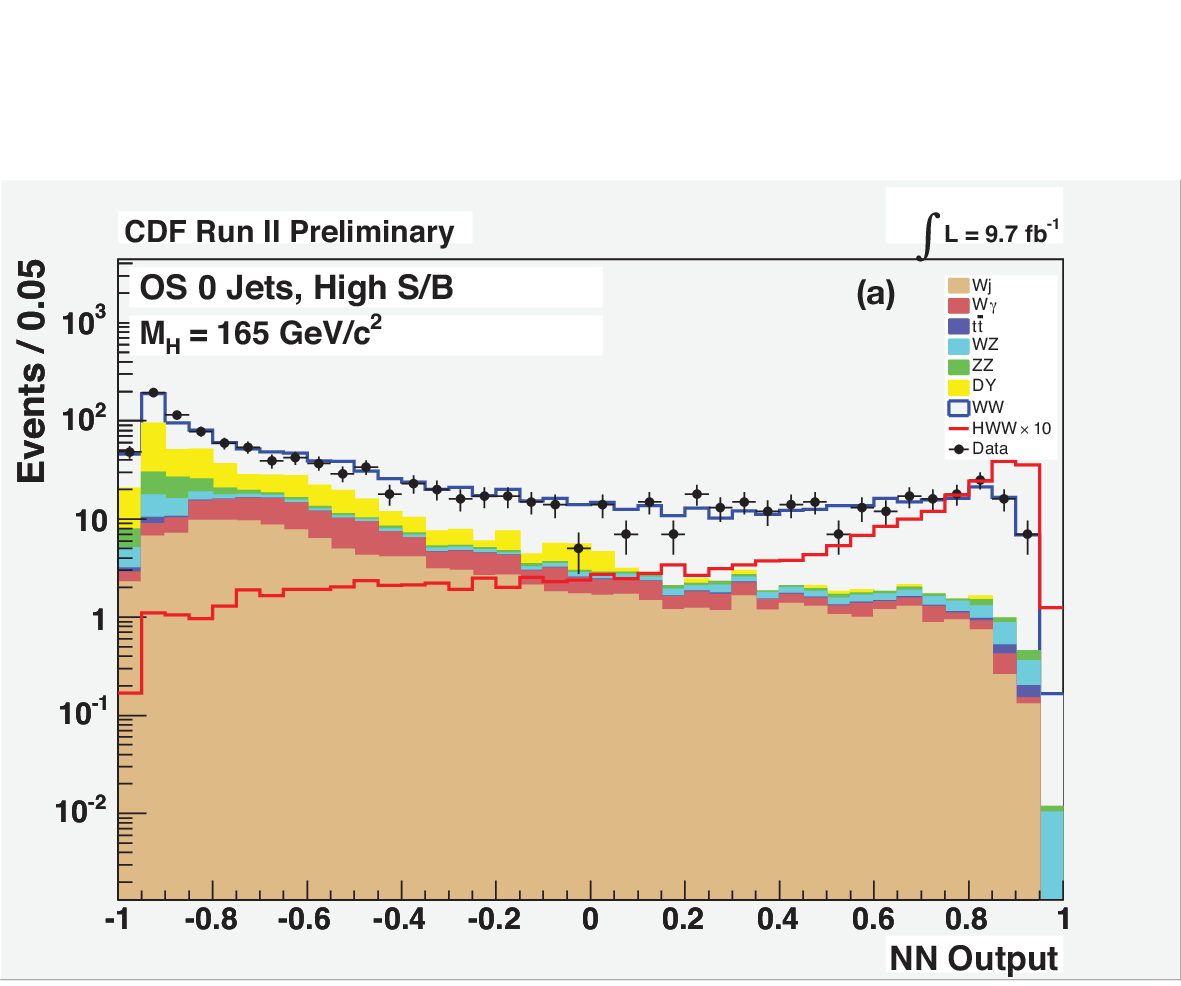} 
\epsfxsize=2.5in\epsfbox{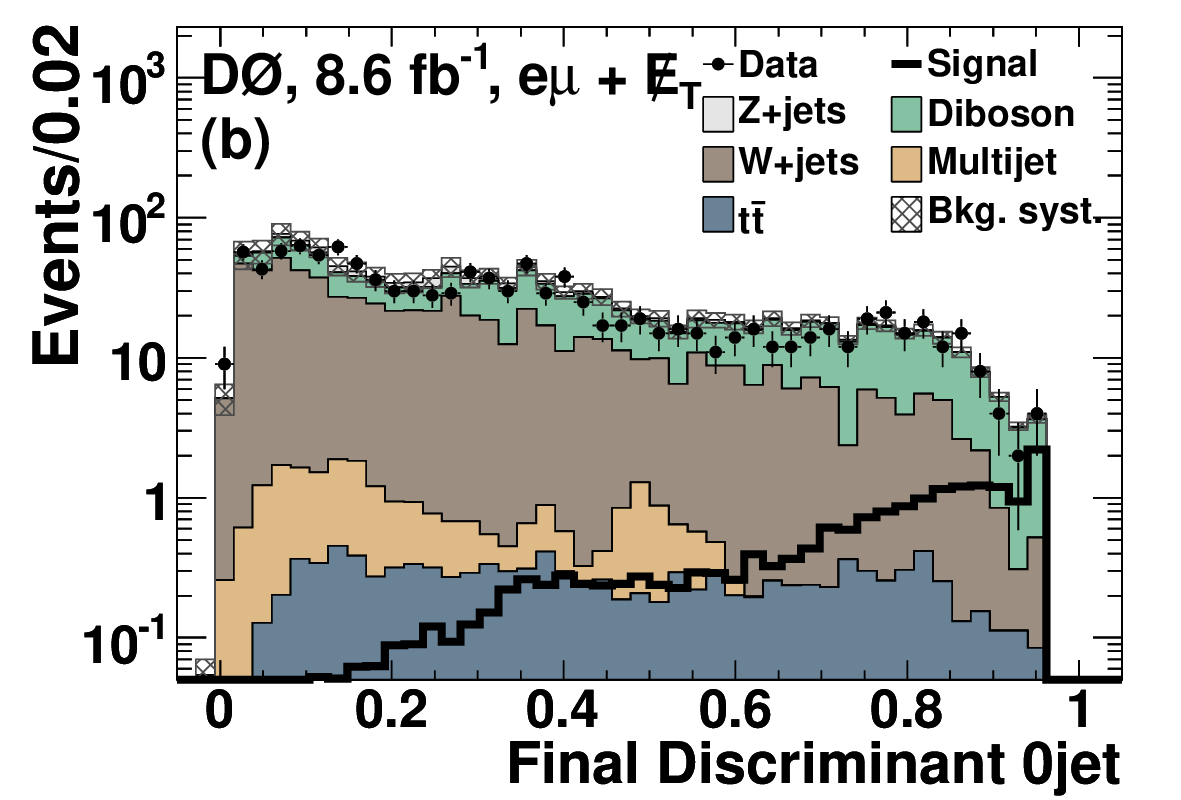} }
\caption[*]{\label{fig:ww_disc}Examples of final discriminant outputs for the search for $H\rightarrow WW$.  Plot (a) shows the distribution of neural network output values from the CDF data compared to the expectation from backgrounds and from a 165 GeV Higgs boson, for events with 0 jets, while (b) shows a similar comparison using D0 events in the $e\mu$ channel.}
\end{center}
\end{figure}

For $m_H > \approx 135$ GeV, the most sensitive search mode at the Tevatron is $H\rightarrow WW \rightarrow \ell \nu \ell^\prime \nu$, where $\ell$ is an electron or muon.  Using this channel, the Tevatron was able to exclude values for $m_H$ near 160 GeV~\cite{TEVexclusion}, marking the first exclusion after LEP2.  Both CDF and D0 have now updated their searches to use the entire Run II data sample.  Figure~\ref{fig:ww_disc} shows the distribution of the final discriminant values in a subset of the data (the analysis is performed in separate sub channels with different signal to background (S/B) ratios to maximize the sensitivity) for both CDF~\cite{CDFWW} and D0~\cite{D0WW}.  The data is in good agreement with the background-only hypothesis, and there is no significant indication of the presence of the Higgs boson in this channel.  

\begin{figure}[!thb]
%\vspace*{4.0cm}
\begin{center}
%\special{psfile=Tev_comb_WW_limit.eps voffset=0 vscale=50
%hscale= 50 hoffset=0 angle=0}
\centerline{\epsfxsize=4.0in\epsfbox{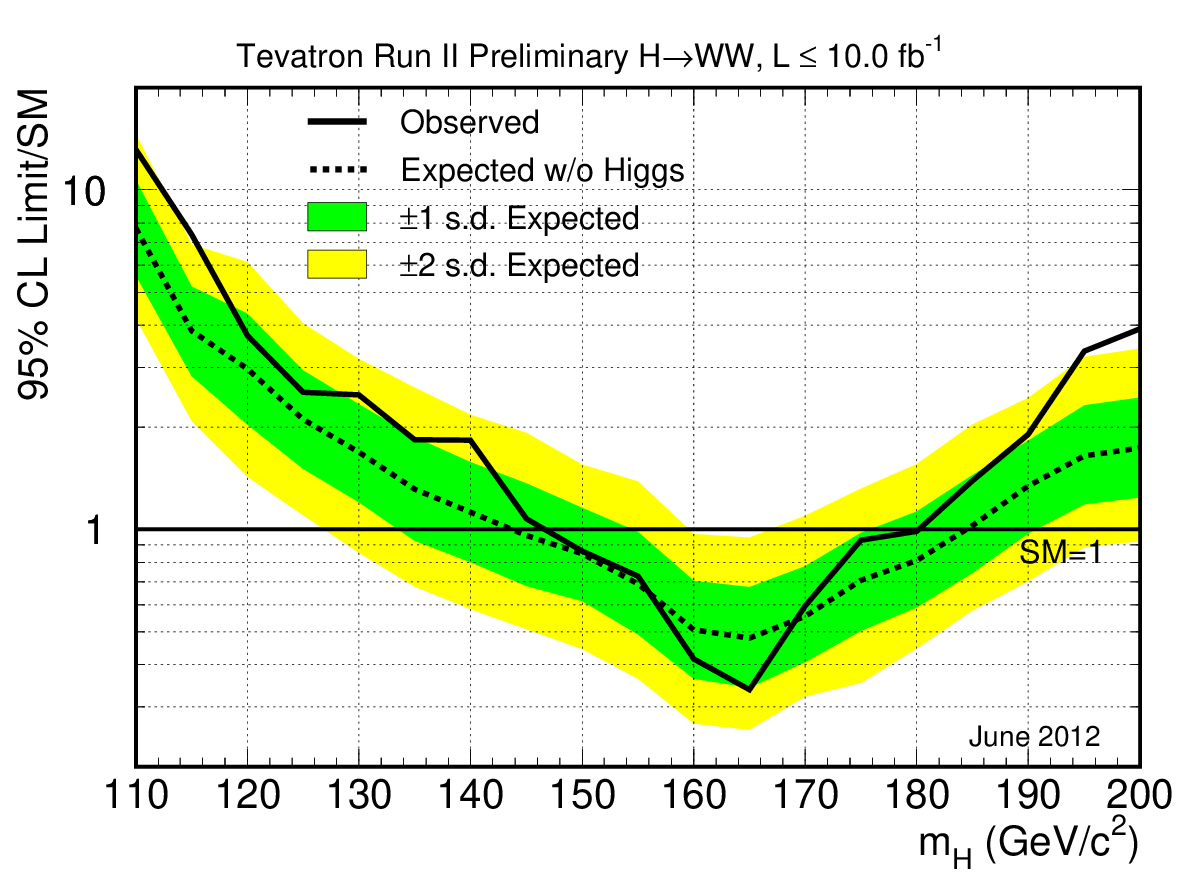}}
\caption[*]{\label{fig:ww_limit}95\% C.L. upper limit on the Higgs boson production cross section as a function of Higgs boson mass, normalized to the SM expectation, as determined using combined CDF and D0 data in the $WW$ decay channel.}
\end{center}
\end{figure}

The combined CDF and D0 data samples are used to set limits on the Higgs production cross section as a function of $m_H$~\cite{TEVcomb}, as shown in Fig.~\ref{fig:ww_limit}.  Higgs boson masses between 147 and 180 GeV are excluded by the Tevatron data.

\subsection{Searches optimized for low mass}

For $m_H < \approx 135$ GeV the dominant decay mode is to $b\bar{b}$, but a search requiring only two $b$ jets in the final state, as expected for $gg \rightarrow H \rightarrow b\bar{b}$ production, would be overwhelmed by multijet background.  Therefore we search for the associated production of the Higgs boson with a $W$ or a $Z$ boson, which comprises only about 10\% of the Higgs boson production cross section, but which can include hard leptons from the decay of the $W$ or $Z$.  Requiring such a lepton in addition to the $b\bar{b}$ pair greatly suppresses background.  This results in three search modes:  $WH \rightarrow \ell\nu b\bar{b}$, $ZH \rightarrow \ell\ell b\bar{b}$, and $ZH \rightarrow \nu\bar{\nu}b\bar{b}$. The S/B ratio at the Tevatron is better than that at the LHC in these channels, allowing the Tevatron results for $H\rightarrow bb$ to remain competitive with those from the LHC.   

Among the three search modes,  $ZH \rightarrow \ell\ell b\bar{b}$ has the cleanest signature and allows the full kinematic reconstruction of the final state, but has the smallest expected yield, $WH \rightarrow \ell\nu b\bar{b}$ has larger yield and background, and $ZH \rightarrow \nu\nu b\bar{b}$ benefits from a larger signal and from the contribution of $WH$ events in which the charged lepton escapes detection, but must overcome a challenging multijet background.

\begin{figure}[!thb]
%\vspace*{4.0cm}
\begin{center}
\centerline{\epsfxsize=3.0in\epsfbox{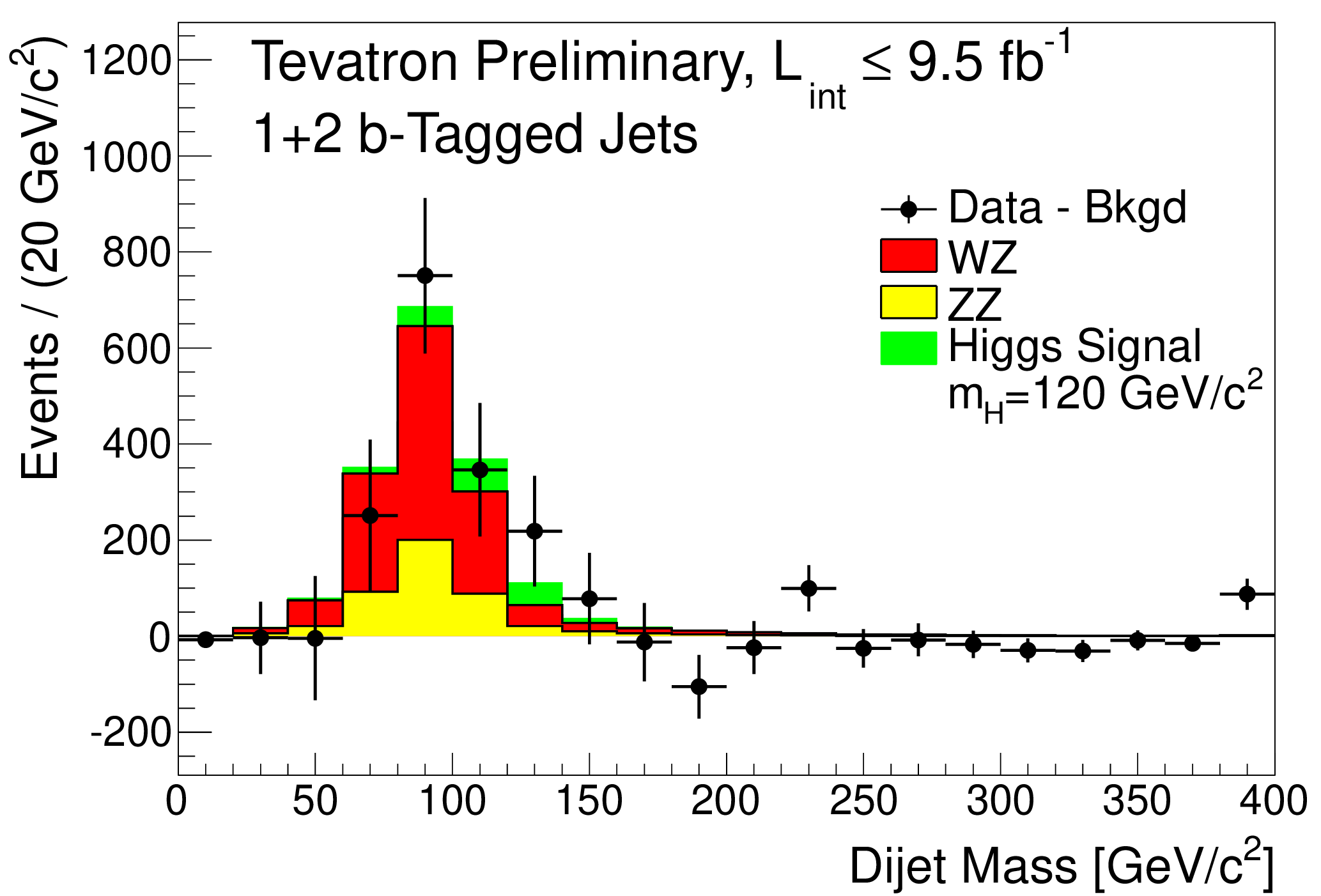}}
\caption[*]{\label{fig:diboson_mass}Background-subtracted dijet mass distribution for events selected using the same tools used for the $WH$ and $ZH$ searches, reconfigured to treat $WZ$ and $ZZ$ as signal.  The plots shows the combination of CDF and D0 data in all search channels}
\end{center}
\end{figure}

Before drawing conclusions about the consistency of the Tevatron data with the presence of a Higgs boson, it is important to verify the analysis techniques by measuring the cross section of a known process that leads to the same final states as has a comparable event yield.  Diboson ($WZ$ and $ZZ$) production where one $Z$ boson decays to $b\bar{b}$, is an excellent verification process, as it results in signatures similar to $WH$ and $ZH$ with the exception that the $b\bar{b}$ mass distribution peaks at the $Z$ pole.  The expected diboson event yield is close to an order of magnitude larger that for SM associated Higgs boson production.  However, backgrounds are also somewhat larger at the $Z$ boson mass than at $m_H$.  The $b\bar{b}$ mass distribution observed in the CDF and D0, with all SM processes other than $WZ$ and $ZZ$ subtracted, is shown in Fig.~\ref{fig:diboson_mass}.  A clear peak is present at the $Z$ mass, with a significance of more than 4.5 standard deviations (s.d.), and a measured cross section of $\sigma_{WZ+ZZ} = 3.9 \pm 0.9$ pb, to be compared to the SM expectation of $4.4 \pm 0.3$ pb.  

\begin{figure}[!thb]
%\vspace*{4.0cm}
\begin{center}
\centerline{\epsfxsize=3.0in\epsfbox{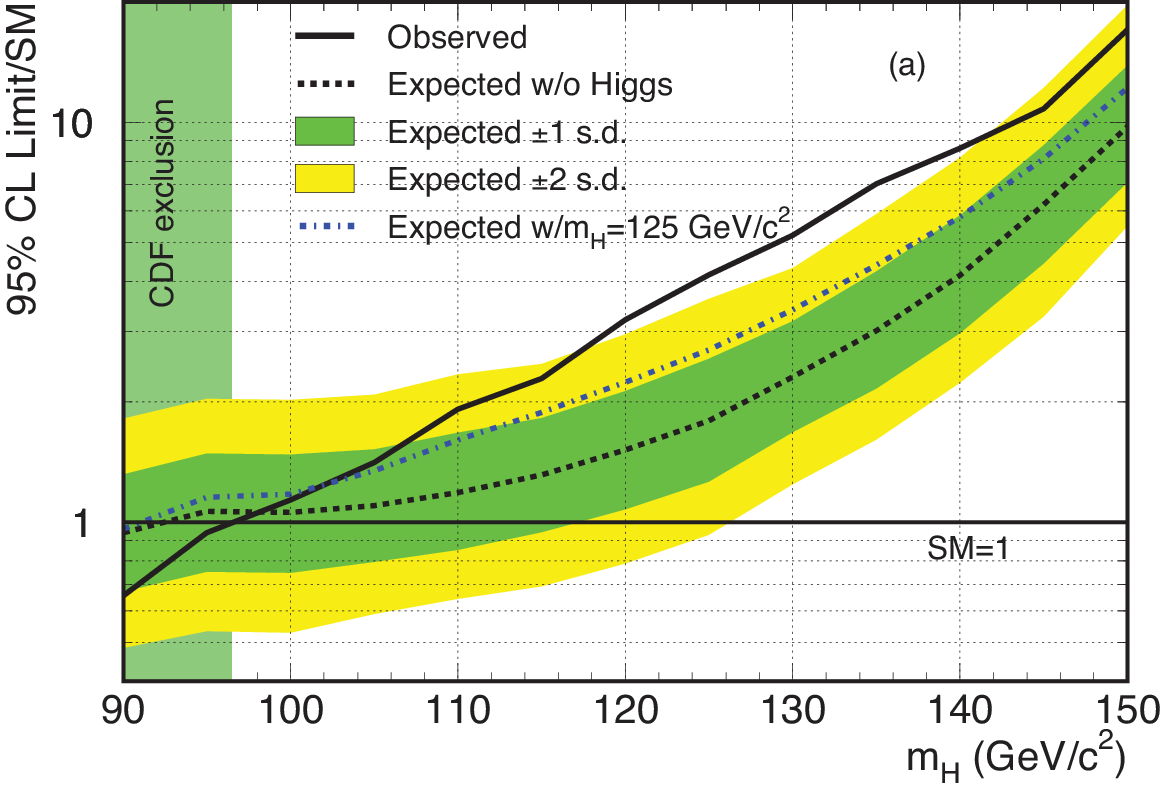}
\epsfxsize=3.0in\epsfbox{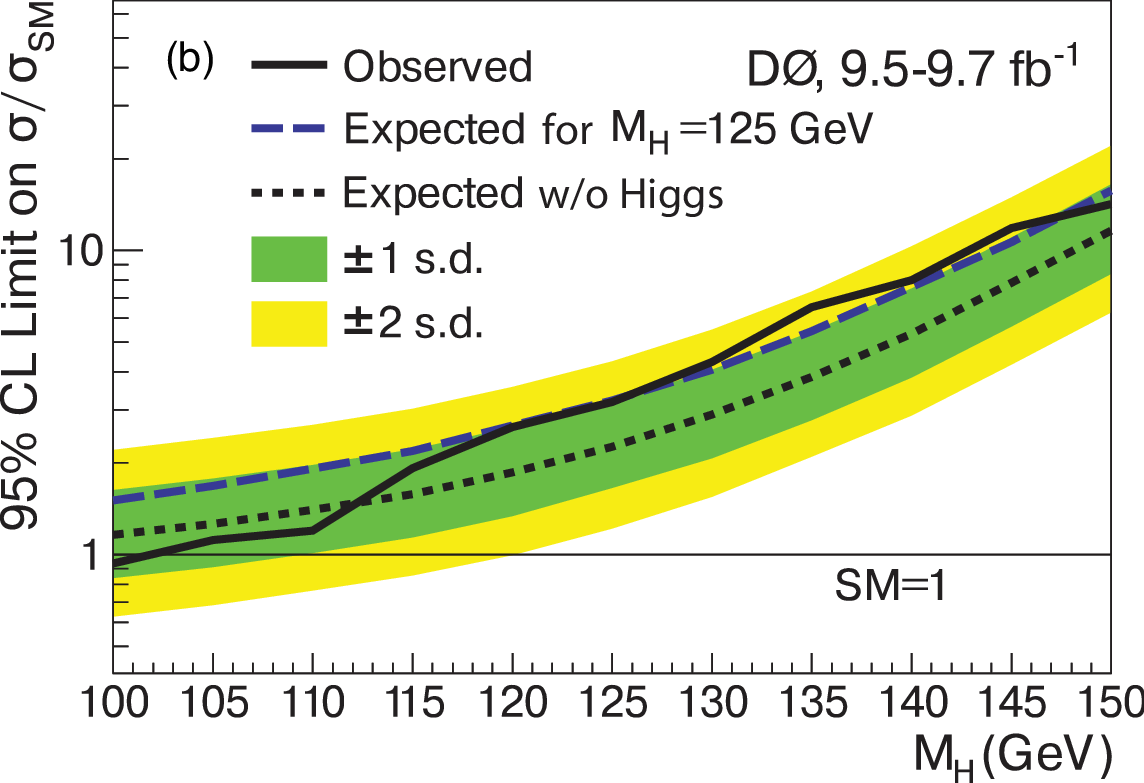}}
\centerline{\epsfxsize=3.5in\epsfbox{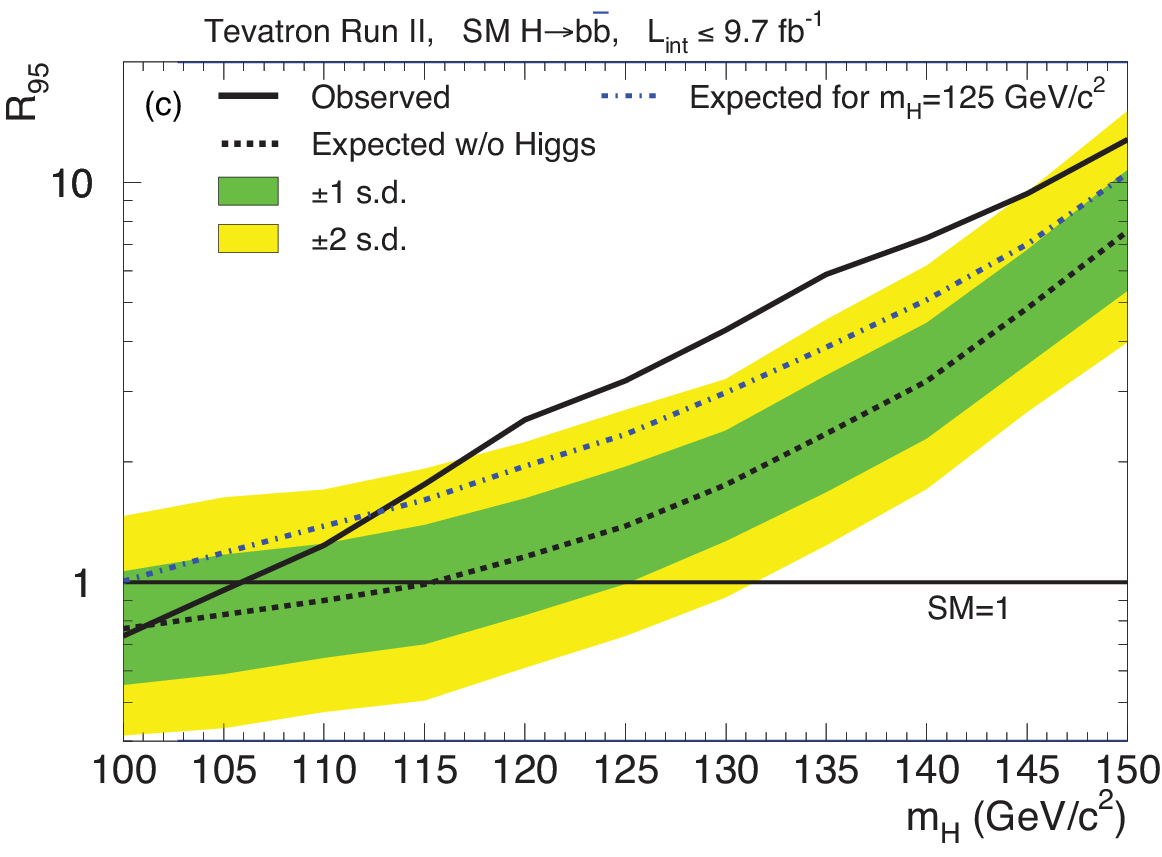}}
\caption[*]{\label{fig:comb_bb_limit}95\% C.L. upper limit on the Higgs boson cross section (normalized to the SM prediction) obtained from the combination of searches in the $VH\rightarrow b\bar{b}$ channels at CDF (a), D0 (b), and the combination of the two (c).}
\end{center}
\end{figure}

\begin{figure}[!thb]
%\vspace*{4.0cm}
\begin{center}
\centerline{\epsfxsize=3.0in\epsfbox{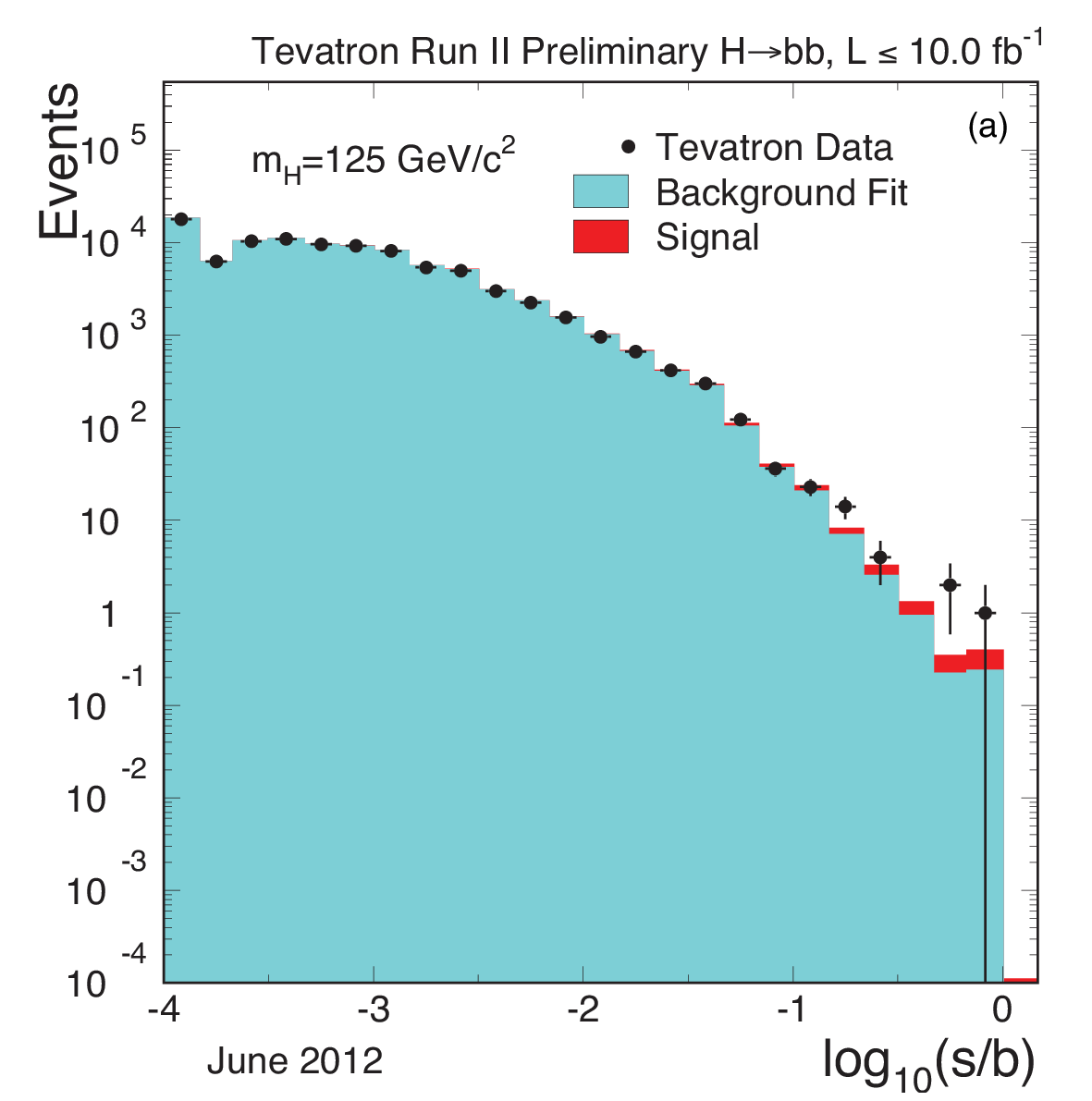}
\epsfxsize=3.0in\epsfbox{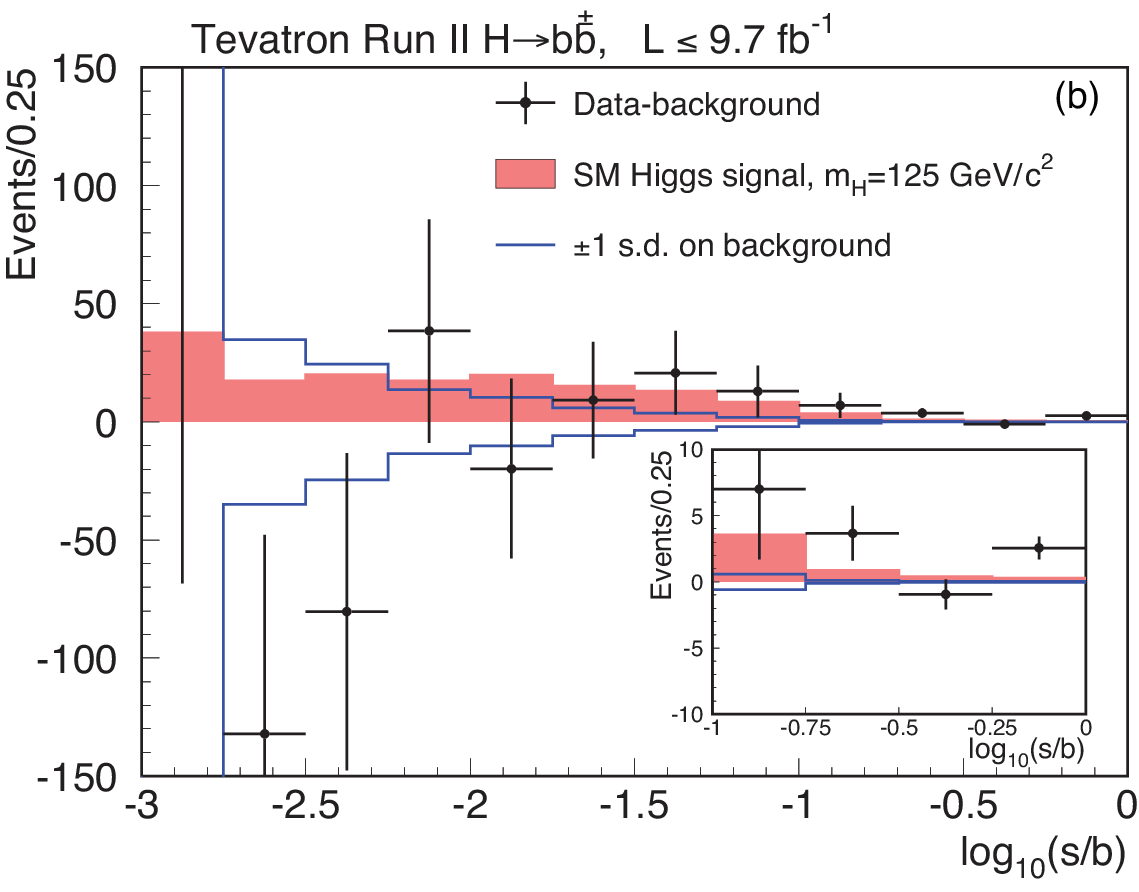}}
\caption[*]{\label{fig:comb_bb_data}(a) Distribution of $VH \rightarrow b\bar{b}$ candidates from CDF and D0 combined, compared to the expectations from background and signal.  Events are binned according to the expected S/B ratio.  (b) The distribution in the highest-S/B bins after subtracting the background.}
\end{center}
\end{figure}

Turning the analysis apparatus to the Higgs boson search, CDF and D0 both find an excess over the SM background predictions in the range $120 < m_H < 145$ GeV~\cite{CDFcomb,D0bb}, as shown in Fig.~\ref{fig:comb_bb_limit}(a) and (b).  These excesses have a significance of 2.5 s.d. for CDF and 1.5 s.d. for D0.  The consistency of the data with the background-only and signal plus background processes can be visualized by combining events from both experiments across all analysis channels~\cite{TEVbb} by binning them according to the S/B ratio implied by their multivariate discriminant value, as shown in
Fig.~\ref{fig:comb_bb_data}. The Higgs boson production cross section limit implied by the data is shown in Fig.~\ref{fig:comb_bb_limit} (c).

\begin{figure}[!thb]
%\vspace*{4.0cm}
\begin{center}
\centerline{\epsfxsize=3.8in\epsfbox{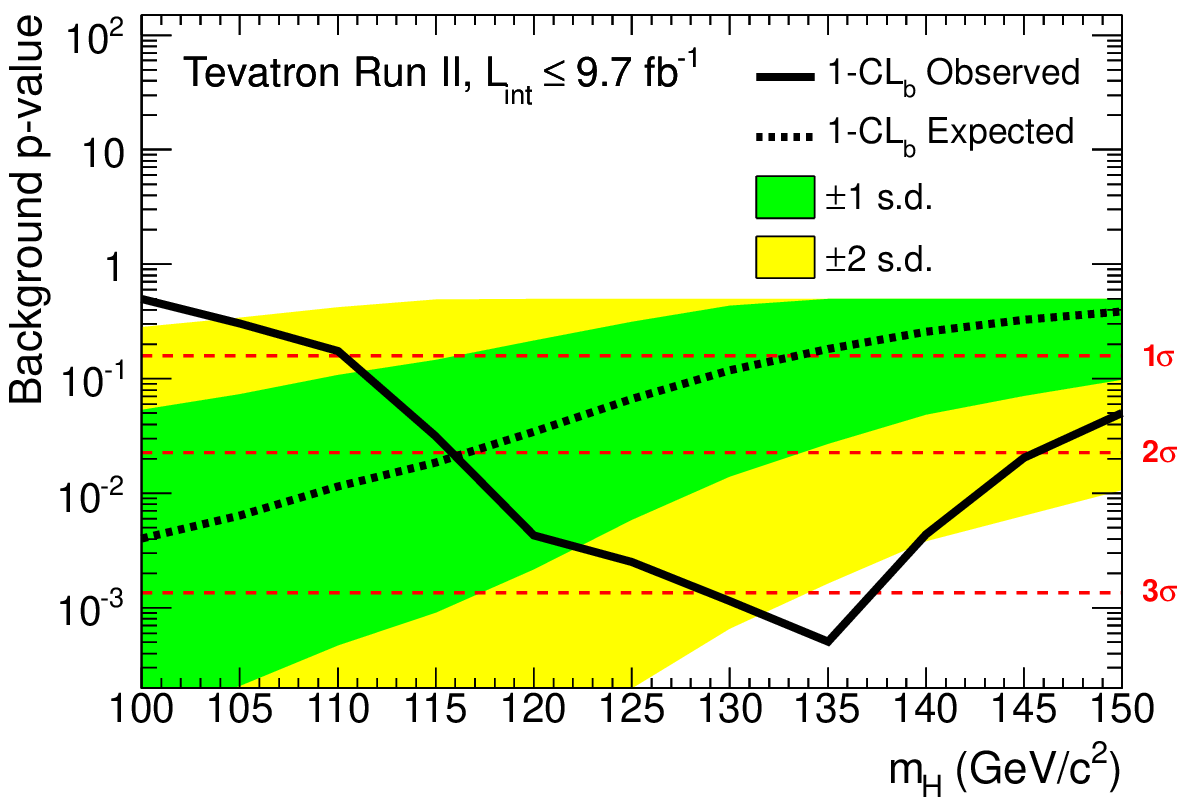}}
\caption[*]{\label{fig:comb_bb_significance}Probability for the background-only hypothesis to describe the combined CDF and D0 data as a function of Higgs boson mass.}
\end{center}
\end{figure}

The significance of the excess can be quantified by calculating the probability for the observed distribution of events to arise from background only, as a function of the $m_H$ value assumed in the multivariate discriminant.  As shown in Fig.~\ref{fig:comb_bb_significance}, this probability falls below $10^{-3}$ near 135 GeV, corresponding to a significance of 3.3 s.d.  Accounting for the look-elsewhere effect brings the global significance to 3.1 s.d.  If we assume that the particle observed at the LHC is in fact the SM Higgs, and therefore that any excess relevant to the Higgs should be calculated at 125 GeV, a significance of 2.8 s.d. is found. This is the most significant evidence to date for Higgs boson decays to fermions.

\begin{figure}[!thb]
%\vspace*{4.0cm}
\begin{center}
\centerline{\epsfxsize=4.0in\epsfbox{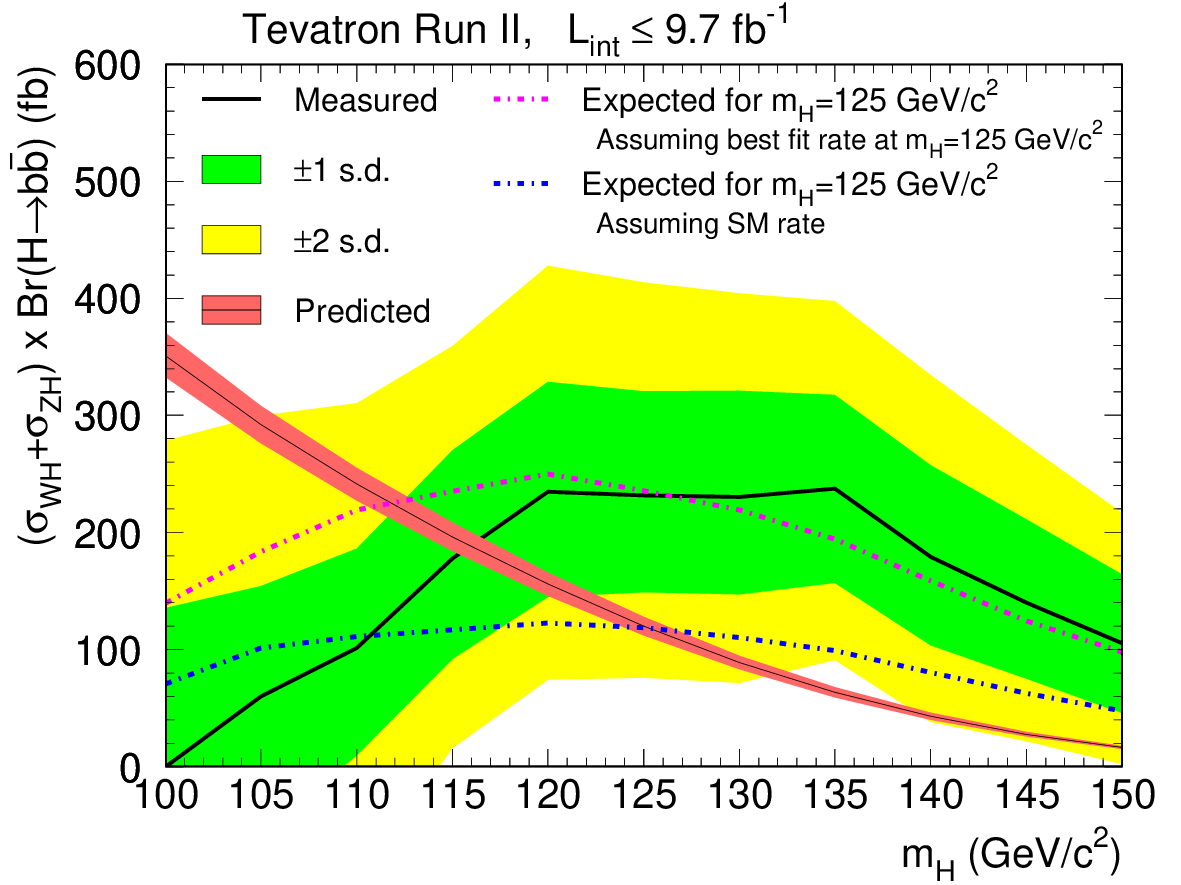}}
\caption[*]{\label{fig:comb_bb_xs}Measured $VH$ production cross section, with one- and two-s.d. error bands, as a function of Higgs boson mass, using the combined CDF and D0 data.  The red band shows the expectation from the SM, with its uncertainty.}
\end{center}
\end{figure}

The measured value of the Higgs boson cross section times $b\bar{b}$ branching ratio is $0.23^{+0.09}_{-0.08}$ pb, larger than (but not inconsistent with) the SM expectation of $0.12 \pm 0.01$ pb, assuming that $m_H = 125$ GeV.  The cross section times branching ratio as a function of $m_H$ is shown in Fig.~\ref{fig:comb_bb_xs}.  

\begin{figure}[!thb]
%\vspace*{4.0cm}
\begin{center}
\centerline{\epsfxsize=3.0in\epsfbox{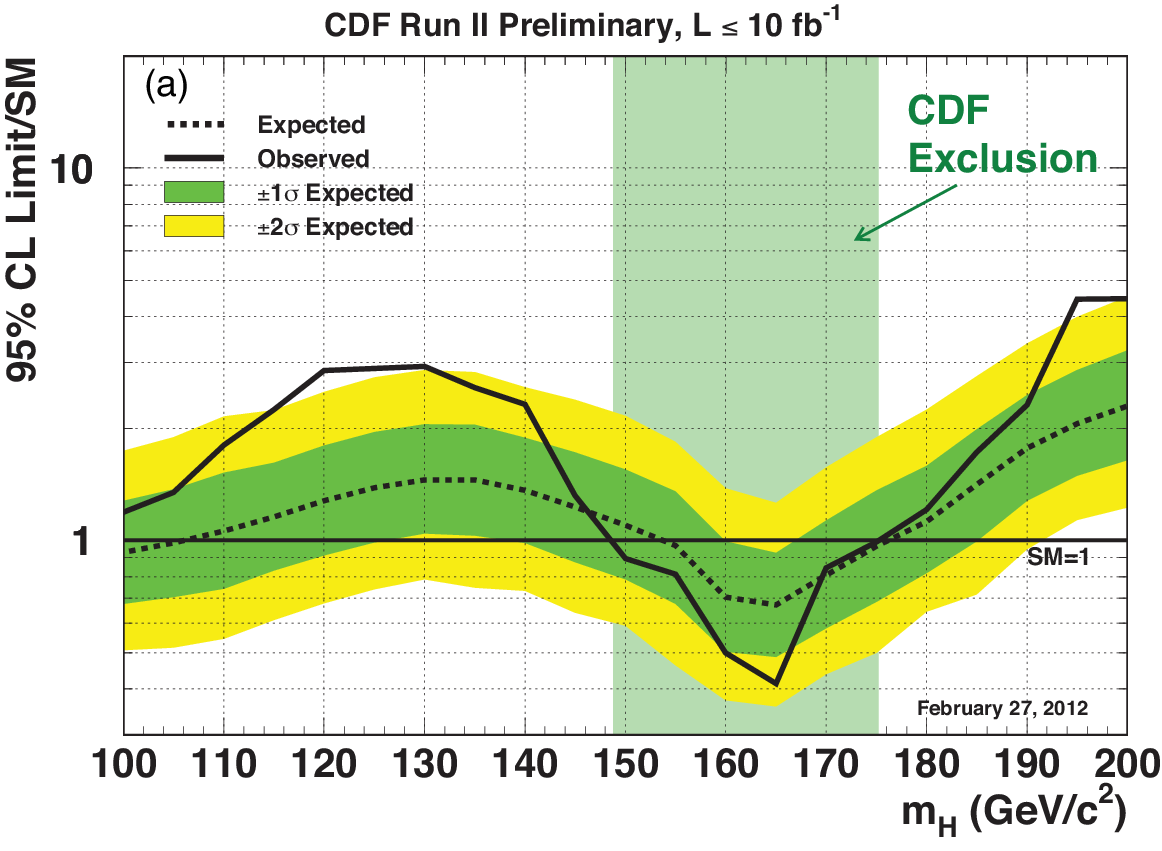}
\epsfxsize=3.0in\epsfbox{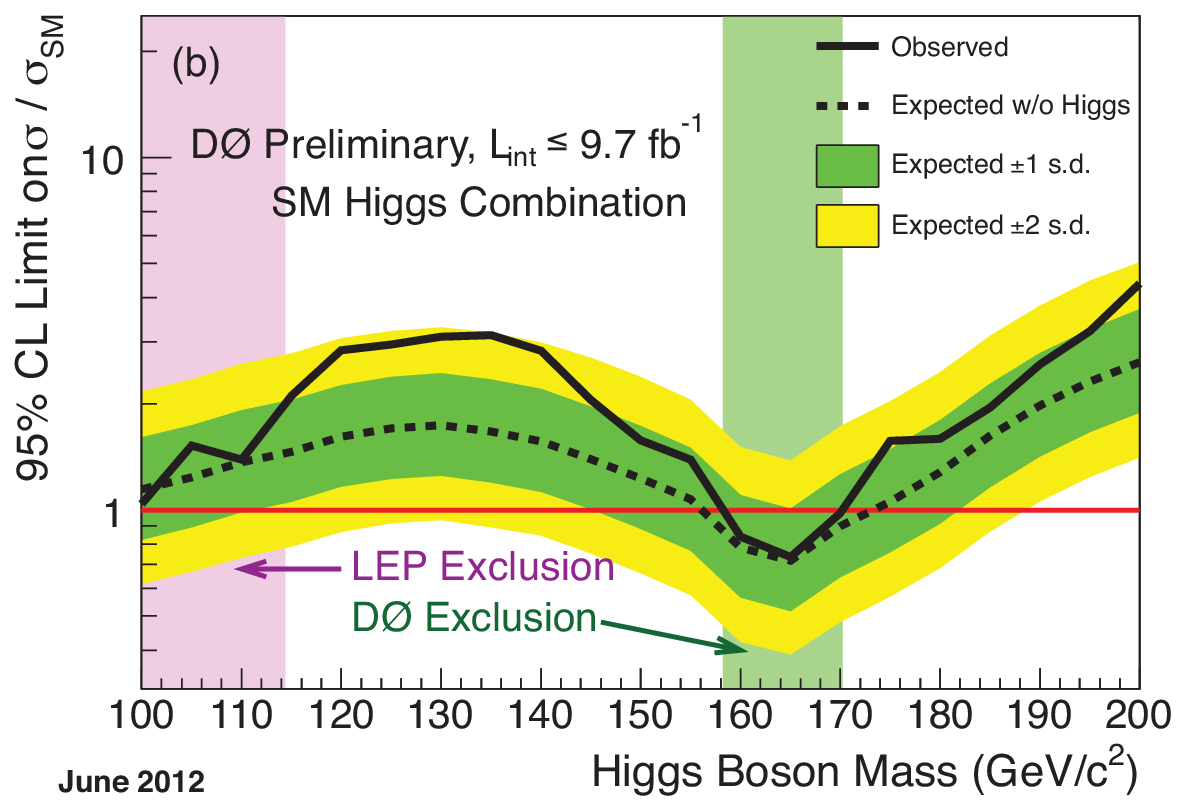}}
\centerline{\epsfxsize=3.5in\epsfbox{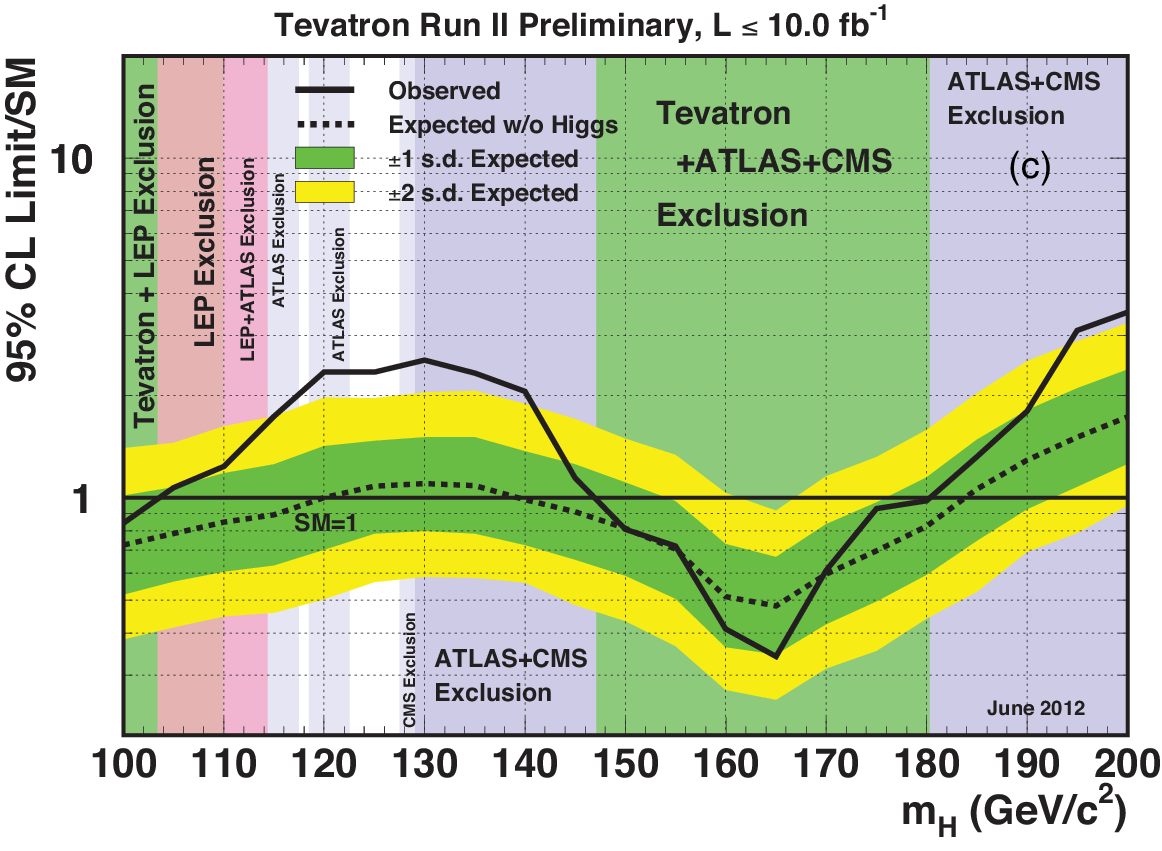}}
\caption[*]{\label{fig:comb_limit}95\% C.L. upper limit on the Higgs boson cross section (normalized to the SM prediction) obtained from the combination of searches in all channels at CDF (a), D0 (b), and the combination of the two (c).}
\end{center}
\end{figure}

\begin{figure}[!thb]
%\vspace*{4.0cm}
\begin{center}
\centerline{\epsfxsize=4.0in\epsfbox{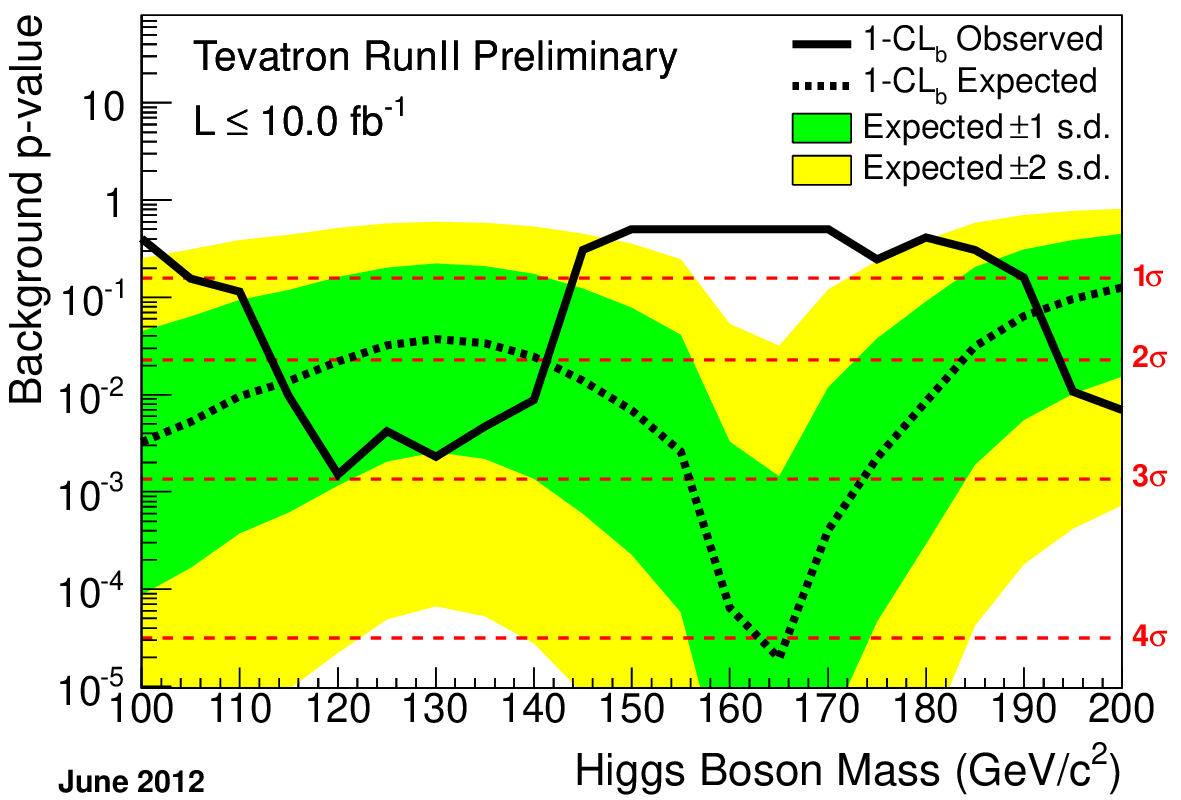}}
\caption[*]{\label{fig:comb_significance}Probability for the background-only hypothesis to describe the combined CDF and D0 data as a function of Higgs boson mass.}
\end{center}
\end{figure}

While the $WW$ and $b\bar{b}$ decay modes provide the greatest sensitivity to the Higgs boson at the Tevatron, searches have also been conducted in other modes, such as $H\rightarrow \gamma\gamma$, $H \rightarrow \tau\tau$, $WH\rightarrow WWW$, and $t\bar{t}H$ production. While none of these channels are sensitive to SM Higgs production individually, including them increases the overall sensitivity at the Tevatron by 10 - 20\%~\cite{CDFcomb,D0comb}, depending on $m_H$.  The result of the combination of CDF and D0 searches in all modes~\cite{TEVcomb} is shown in Fig.~\ref{fig:comb_limit}.  A SM Higgs boson with mass between 147 and 180 GeV is excluded at 95\% C.L., and at lower masses the limits are less restrictive than expected in the background-only scenario, indicating an excess of events.  The significance of the excess, shown in Fig.~\ref{fig:comb_significance} is 2.5 s.d. (including look-elsewhere effect), driven largely by the $b\bar{b}$ decay mode.

\section{Searches for the Higgs boson in extensions to the SM}

In addition to searching for the Higgs boson in the context of the SM, the CDF and D0 collaborations carried out searches in non-SM scenarios.  In some cases these searches involved a reinterpretation of the results of the SM searches (as the expected production cross section and decay branching fractions would be modified by the new physics), while in other cases they involved searches for novel signatures.

\begin{figure}[!thb]
%\vspace*{4.0cm}
\begin{center}
\centerline{\epsfxsize=4.0in\epsfbox{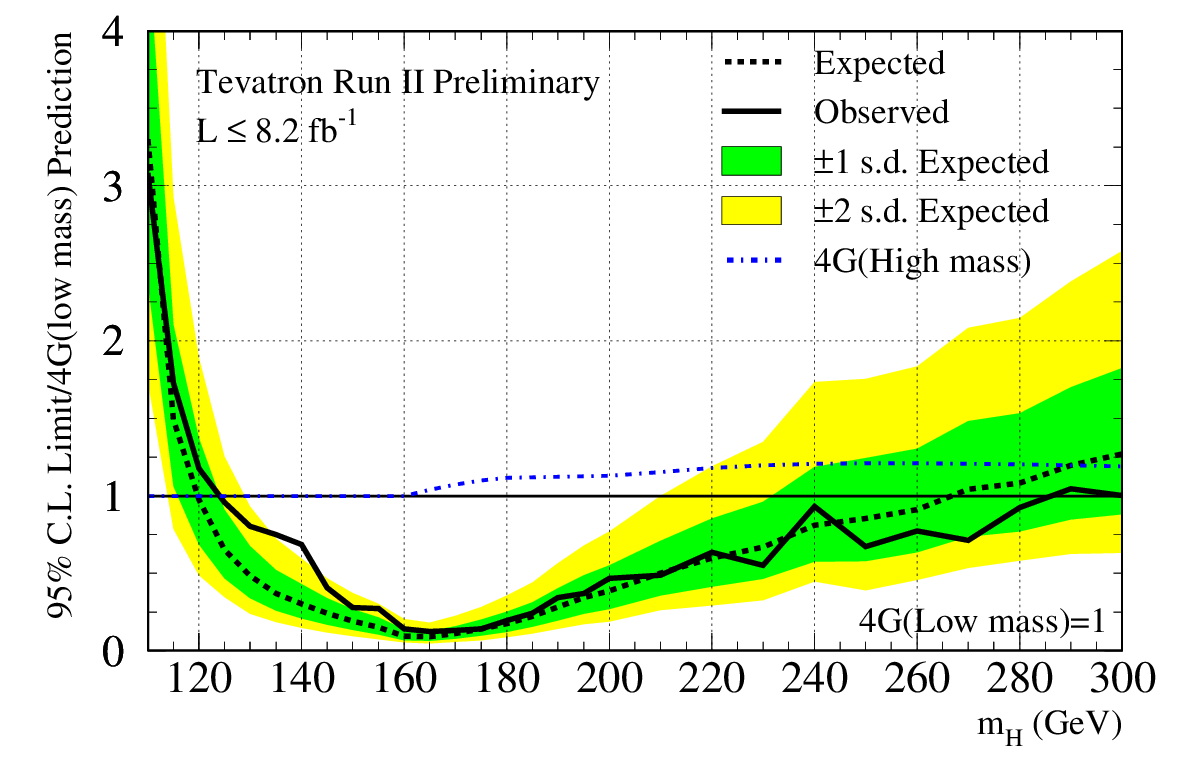}}
\caption[*]{\label{fig:comb_4thgen}95\% C.L. upper limit on Higgs boson production under the assumption of the existence of a fourth generation of chiral quarks, using combined CDF and D0 data.}
\end{center}
\end{figure}

One example of the former category is the Higgs search under the assumption of the existence of a fourth generation of quarks.  In this scenario, there would be additional fermion loops contributing to the $gg \rightarrow H$ cross section.  Since the fourth generation quarks must be more massive than the top, each additional quark  loop contributes an amplitude similar to that of a top quark, resulting in a production cross section $\sim9$ times larger than the SM prediction.  Thus the upper limit on the $gg \rightarrow H$ production cross section can constrain the existence of a fourth quark generation.  Under the assumption that the fourth-generation leptons have masses that just exceed the current limits, a fourth quark generation is excluded by the combined CDF and D0 data for $124 < m_H < 286$ GeV~\cite{TEV4thGen}, as shown in Fig.~\ref{fig:comb_4thgen}.  This implies that if the new particle seen at 125 GeV at the LHC is in fact the SM Higgs boson, then a fourth chiral quark generation is excluded.

\begin{figure}[!thb]
%\vspace*{4.0cm}
\begin{center}
\centerline{\epsfxsize=4.0in\epsfbox{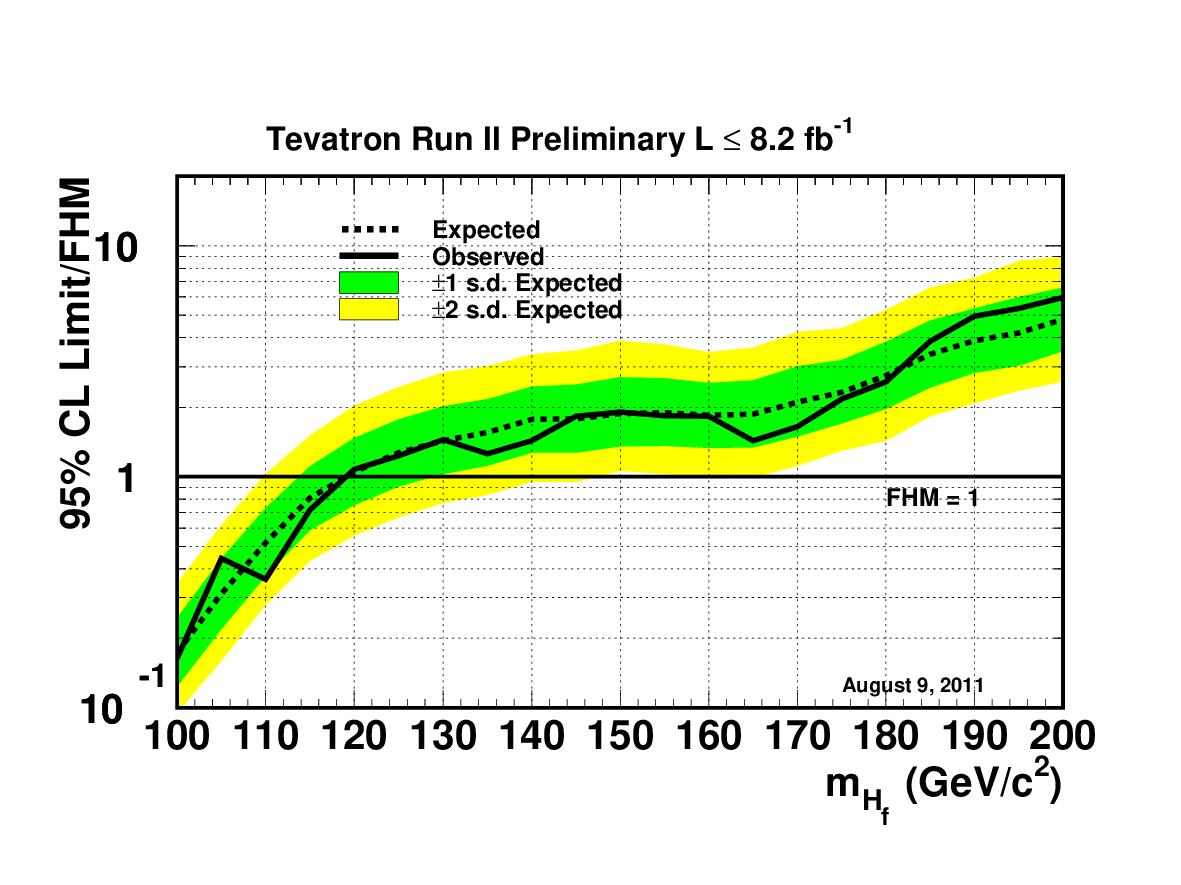}}
\caption[*]{\label{fig:comb_fermiophobic}95\% C.L. upper limit on Higgs boson production under the assumption that the Higgs does not couple to fermions, using combined CDF and D0 data.}
\end{center}
\end{figure}

 One can can also search for deviations from the SM Higgs couplings. One example of this is the search for a fermiophobic Higgs.  In this scenario the gluon fusion production mechanism is greatly suppressed (since Higgs bosons cannot be created via a fermion loop), and the branching fractions to $WW$, $ZZ$, and $\gamma\gamma$ are enhanced (particularly at low $m_H$), since there is no possibility of $H\rightarrow b\bar{b}$.  CDF and D0 have both performed searches optimized for sensitivity to such fermiophobic Higgs bosons in the $WW$ and $\gamma\gamma$ channels, and find no evidence for their production.  The resulting limits on the cross section in the fermiophobic model obtained from the combined CDF and D0 data~\cite{TEVfermiophobic} are shown in Fig.~\ref{fig:comb_fermiophobic}. Fermiophobic Higgs masses below 119 GeV are excluded at 95\% C.L.

In supersymmetry the Higgs sector is extended to include at least two complex doublets, resulting in five physical bosons (three neutral, two charged) after electroweak symmetry breaking.  Under the Minimal Supersymmetric Model (MSSM)~\cite{MSSM} the production of neutral Higgs bosons (here generically referred to as $\phi$) is proportional to $\tan^2\beta$. The dominant production mode is  $p\bar{p} \rightarrow \phi b$ modes, and the dominant decay of the $\phi$ is to $b\bar{b}$, resulting in a $bbb$ final state.  The main complication of a search in the $bbb$ mode is understanding the large multijet background, which is difficult to simulate and subject to large theoretical uncertainties.  This background is modeled using a data-driven approach that considers the distribution of jet transverse momenta and the number of tagged $b$ jets in each event.  

\begin{figure}[!thb]
%\vspace*{4.0cm}
\begin{center}
\centerline{\epsfxsize=3.0in\epsfbox{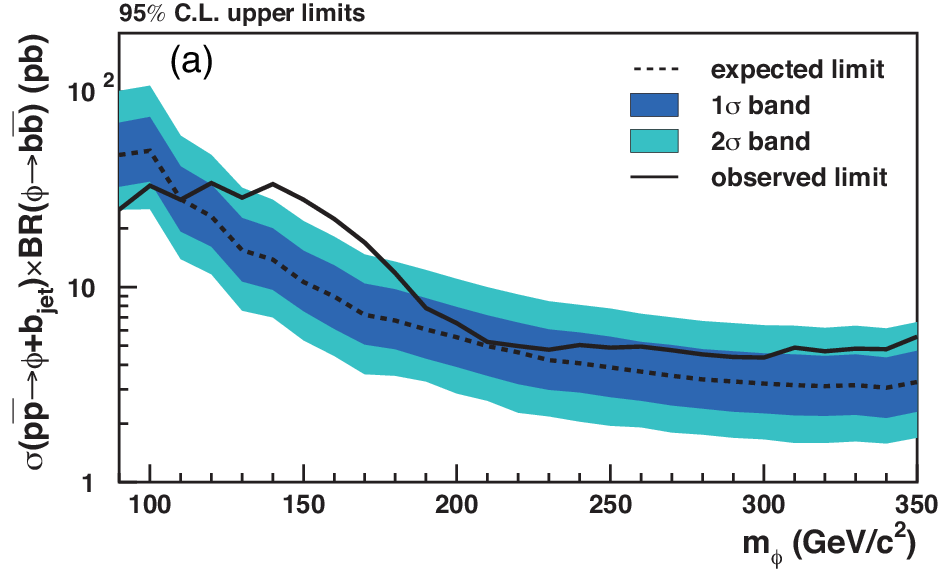}
\epsfxsize=3.0in\epsfbox{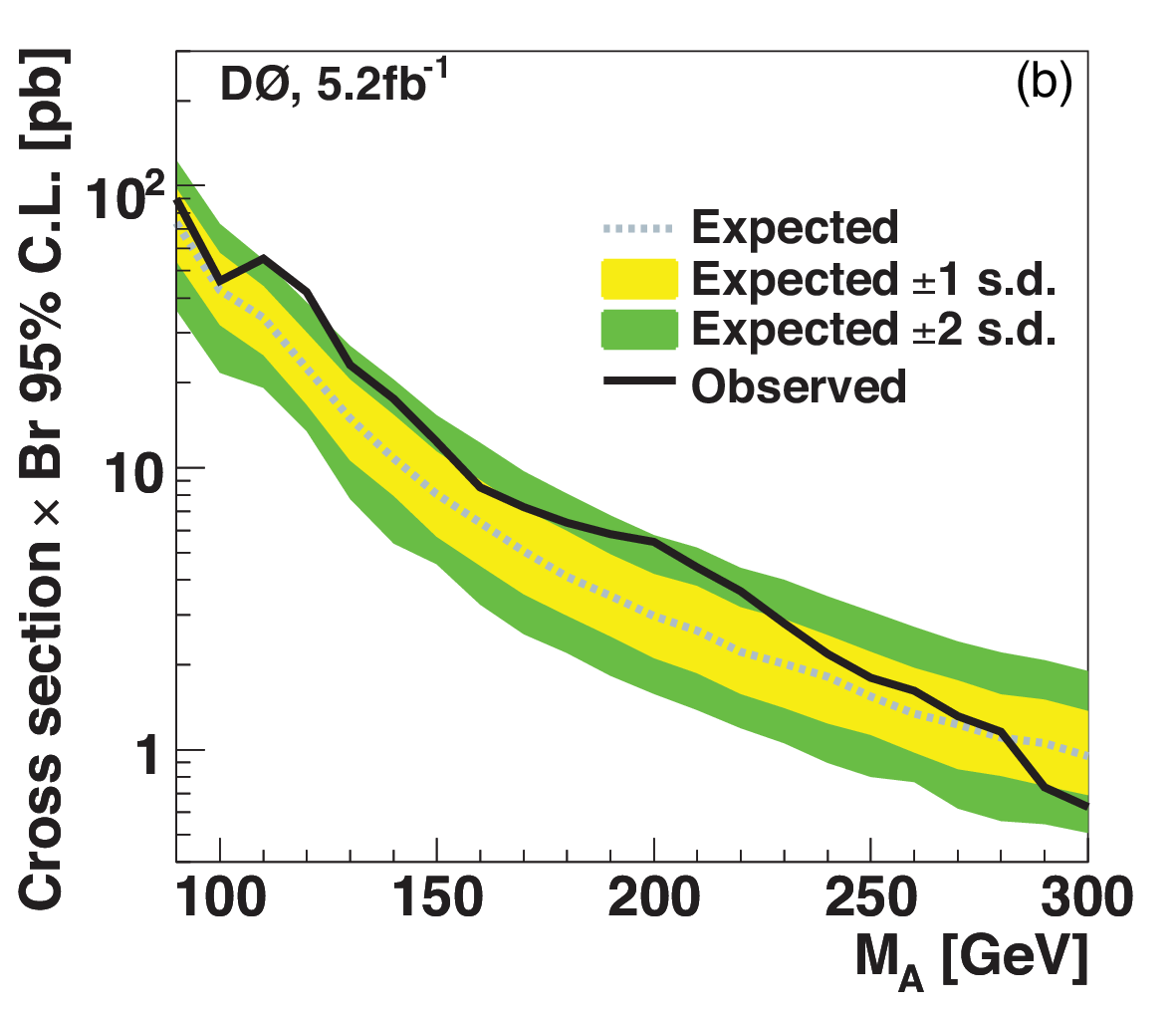}}
\centerline{\epsfxsize=3.5in\epsfbox{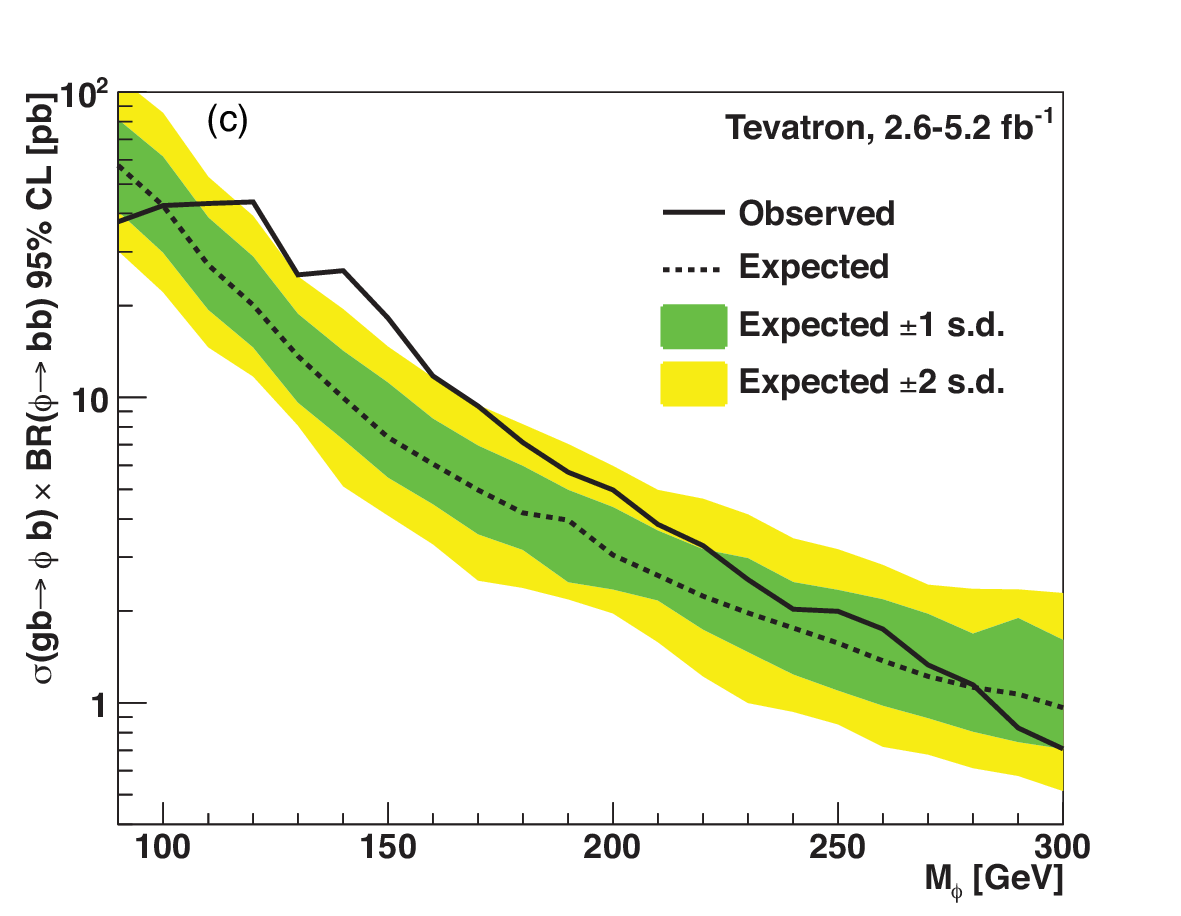}}
\caption[*]{\label{fig:comb_bbb_limit}95\% C.L. upper limit on the Higgs boson cross section (normalized to the SM prediction) obtained from the combination of searches in all channels at CDF (a), D0 (b), and the combination of the two (c).}
\end{center}
\end{figure}

The results are presented in Fig.~\ref{fig:comb_bbb_limit}.  The CDF data sample has an excess of events over background at the 2.8 s.d. level for a $\phi$ mass of 120 GeV~\cite{CDFbbb}, while D0's largest excess is at the 2.5 s.d. level for $m_\phi = 150$ GeV~\cite{D0bbb}.  In both cases accounting for the look-elsewhere effect reduces the overall significance to 2 s.d.  The combined limits~\cite{TEVbbb} are also shown in Fig.~\ref{fig:comb_bbb_limit}.  The fact that the CDF and D0 excesses are at different mass hypotheses results in a combined excess at the 2 s.d. level in the $120 < m_\phi < 150$ GeV region, no larger than the excess seen in the individual experiments.  

\begin{figure}[!thb]
%\vspace*{4.0cm}
\begin{center}
\centerline{\epsfxsize=4.0in\epsfbox{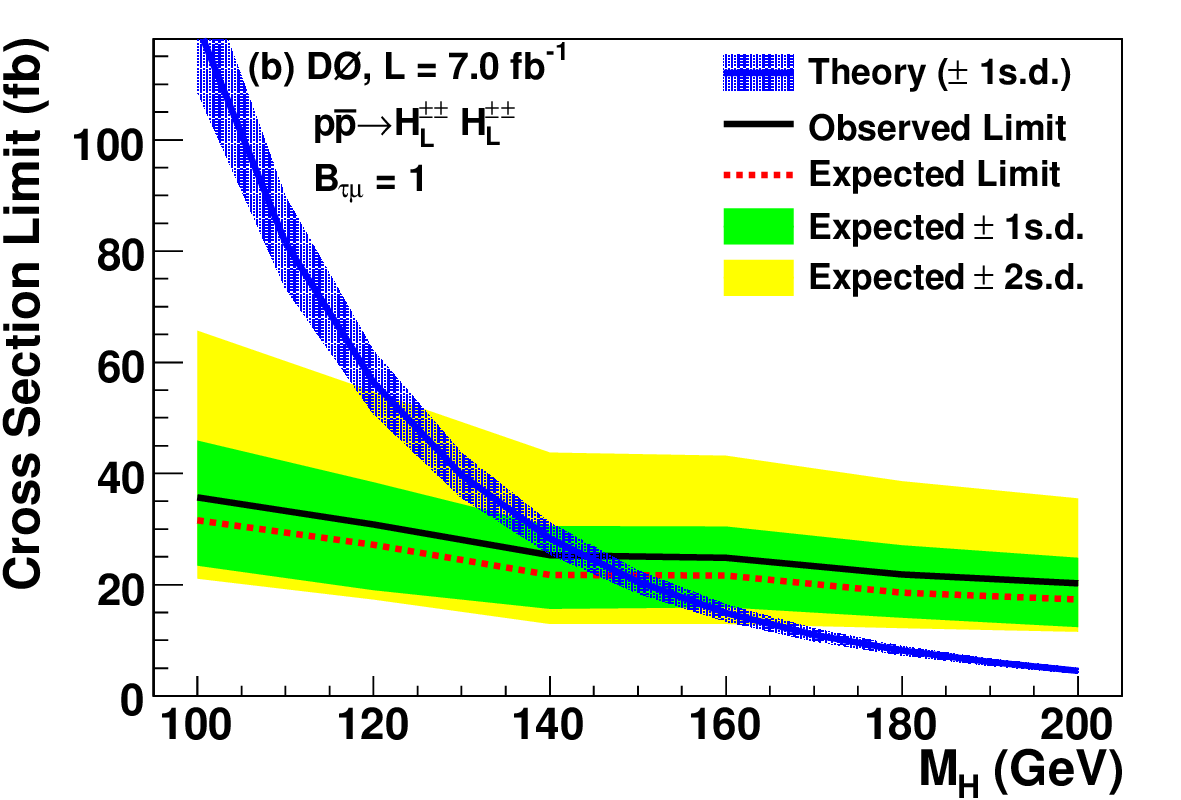}}
\caption[*]{\label{fig:d0_hpp_limit}95\% C.L. upper limit on doubly-charged Higgs boson production as a function of Higgs boson mass obtained using D0 data under the assumption that the branching fraction to $\tau\tau$ is 100\%.}
\end{center}
\end{figure}

Some extensions to the SM, including left-right symmetric~\cite{LRsym} and Little Higgs~\cite{littlehiggs} models, predict the existence of doubly-charged Higgs bosons.  D0 searches for these particles using the mode $p\bar{p} \rightarrow \ell^\pm\ell^\pm\tau^\mp\tau^\mp$ where the taus decay hadronically, representing the first search for the tau pair decay of doubly-charged Higgs bosons~\cite{D0H++}.  
The data is consistent with expectations from the SM, and the resulting lower limits on the mass of a
doubly-charged Higgs boson depend upon the specific model being considered.  An example, under the assumption that the $H^{\pm\pm}$ decays only to tau pairs, is shown in Fig.~\ref{fig:d0_hpp_limit}.

\begin{figure}[!thb]
%\vspace*{4.0cm}
\begin{center}
\centerline{\epsfxsize=4.0in\epsfbox{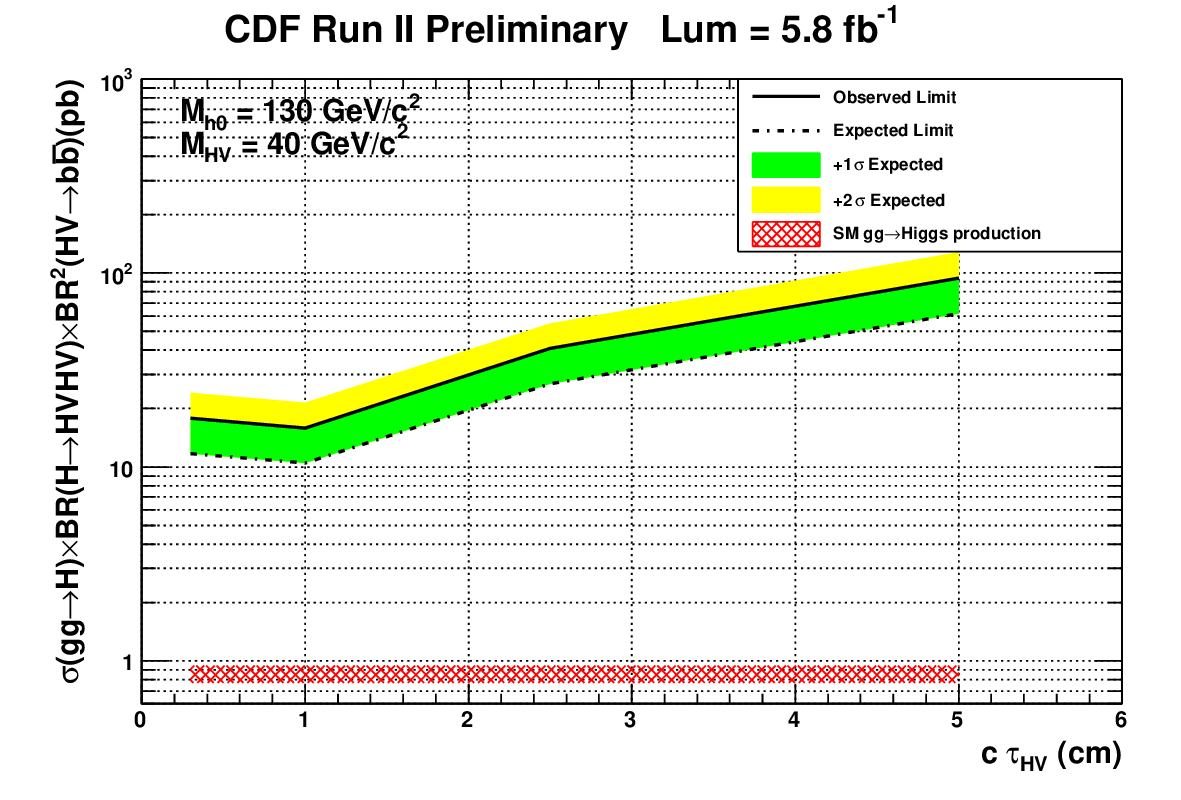}}
\caption[*]{\label{fig:CDF_HV_4b_limit}95\% C.L. upper limit on Higgs decays to pairs of hidden valley particles, each of which subsequently decays to $b\bar{b}$, using CDF data.}
\end{center}
\end{figure}

Another class of extensions to the SM, the ``Hidden Valley'' models~\cite{HV}, posits the existence of a new gauge sector, which is not directly observable. However there are new ``messenger'' particles that couple to both the exotic and SM gauge sectors, and the coupling of these particles to the Higgs boson can result in distinctive signatures.  CDF has searched for Higgs bosons decays under two Hidden Valley scenarios.  In the first of these, the messenger particles are long-lived and decay predominantly to $b$ quark pairs.  Thus Higgs decay to a pair of messenger particles results in two pairs of $b$ jets, each originating form a vertex that is typically centimeters away from the primary interaction vertex.  The CDF data does not indicate an excess for this signature over the SM expectations~\cite{CDFHV}, resulting in the limit on production cross section times branching fraction as a function of messenger particle lifetime shown in Fig.~\ref{fig:CDF_HV_4b_limit}.
In another scenario the Higgs boson decay initiates a cascade of decays into hidden sector particles, the visible trace of which is a large number of low-$p_T$ leptons~\cite{HV_leptons}.  CDF searches for such decays using associated $WH$ and $ZH$ production, where the hard leptons from the $W$ or $Z$ decay are used to trigger the event and suppress background, and  the multiplicity of additional soft leptons in the event is investigated.  No indication of events with anomalously large lepton multiplicity is observed~\cite{CDFHV2}.

\section{Summary}

The CDF and D0 experiments at the Tevatron have searched for the Higgs boson, both in the context of the Standard Model and in extensions upon it.  The strongest evidence for Higgs boson production is found in the search for $H\rightarrow bb$, where an excess over the background expectation in the region $120 < m_H < 150$ GeV is observed, with a significance of 3 s.d. This is the strongest direct evidence of fermonic couplings of the Higgs boson to date.  There is also a more modest excess at low masses in the search for $H\rightarrow WW$.  Searches for non-standard Higgs bosons have also been performed, with no significant deviations from the SM observed.  Though many of the searches use the full Tevatron data sample, updates using refined analysis techniques are still forthcoming. 

\section*{Acknowledgements} 
The work reported here was supported by the DOE and NSF (USA); CONICET and UBACyT (Argentina); ARC (Australia);
CNPq, FAPERJ, FAPESP and FUNDUNESP (Brazil); CRC Program and NSERC (Canada); CAS, CNSF, and NSC
(China); Colciencias (Colombia); MSMT and GACR (Czech Republic); Academy of Finland (Finland); CEA and
CNRS/IN2P3 (France); BMBF and DFG (Germany); INFN (Italy); DAE and DST (India); SFI (Ireland); Ministry
of Education, Culture, Sports, Science and Technology (Japan); KRF, KOSEF and World Class University Program
(Korea); CONACyT (Mexico); FOM (The Netherlands); FASI, Rosatom and RFBR (Russia); Slovak R\&D Agency
(Slovakia); Ministerio de Ciencia e Innovaci\'{o}n, and Programa Consolider-Ingenio 2010 (Spain); The Swedish Research
Council (Sweden); Swiss National Science Foundation (Switzerland); STFC and the Royal Society (United Kingdom);
and the A.P. Sloan Foundation (USA).

\end{document}